\documentclass[[UTF8]{revtex4-1}
\usepackage[utf8]{inputenc}
\usepackage{graphicx}
\usepackage{dcolumn}
\usepackage{bm}
\usepackage{amsmath}
\usepackage{amssymb}
\usepackage{latexsym}
\usepackage{epsfig}
\usepackage{amsbsy}
\usepackage{array}
\usepackage{setspace}
\usepackage{mathrsfs}
\usepackage{geometry}
\usepackage{caption}
\usepackage{multirow}

\begin{document}

\thispagestyle{empty}
	
\title{The non-relativistic expansion of Dirac-Coulomb energy and the non-retarded Breit interaction correction up to $\alpha^8$order}
\author{Wanping Zhou}
\affiliation{School of Physics and Telecommunications, Huanggang Normal University, Huanggang, China, 438000}

\author{Sanjiang Yang}
\affiliation{College of Physics and Electronic Science, Hubei Normal University, Huangshi, China,  435002}

\author{Haoxue Qiao \footnote{Email: qhx@whu.edu.cn}}
\affiliation{School of Physics and Technology, Wuhan University, Wuhan, 430000 China}

\begin{abstract}
The relativistic corrections for the Dirac-Coulomb system are derived through the method of non-relativistic expansion. By expanding the large and small components of the Dirac wave function and the energy eigenvalues in terms of the square of the fine-structure constant $\alpha^2$, we obtain iterative equations for calculating the higher-order relativistic corrections of Coulomb systems. For a single-electron system, the operator results of the iterative equations are consistent with those in the literature Ref[J.Phys.B,At.Mol.Opt.Phys.{\bf 56} 045001]. Using these iterative equations, we numerically calculate the relativistic corrections up to the order of $\alpha^{20}$ for the hydrogen atom, which converge rapidly to the analytical results of the hydrogen atom. For the two-electron Dirac-Coulomb system, we also present iterative equations for calculating high-order energy corrections, as well as numerical energy corrections of ground state up to the order of $\alpha^8$. This work also presents the non-relativistic expansion form of non-retarded Breit interaction correction. The $\alpha^4$ order correction to the Dirac Coulomb energy and non-retarded Breit interaction corresponds precisely to the $\alpha^4$ order relativistic correction. Higher-order expansion terms contribute at even powers of $\alpha$, which represent the contributions from all Coulomb photons and single transverse photons under the non-retarded approximation.
\end{abstract}


\maketitle

\newpage

\section{INTRODUCTION}
In low-charge-$Z$ atomic and molecular systems, the motion of electrons and nuclei is much slower than the speed of light, and the binding energy of the system is much smaller than the electron's rest energy. In hydrogen-like atomic systems, the ratio of the electron's average velocity to the speed of light is $Z\alpha << 1$, and the ratio of the binding energy to the electron's rest energy is $(Z\alpha)^2 << 1$. Therefore, the main characteristics of these systems can be captured by solving the non-relativistic Schrödinger equation. However, for the fine and hyperfine structures of atoms and molecules, higher-order relativistic corrections, quantum electrodynamic corrections, nuclear recoil effects, corrections due to the electromagnetic structure of the nucleus, and non-electromagnetic interactions must be taken into account. Non-relativistic quantum electrodynamics(NRQED) is an effective theory to study the above corrections \cite{CASWELL1986437,PhysRevD.53.4909,PhysRevD.55.7267,JPhysB.31.5123,PhysRevA.72.062102}. The NRQED effective Hamiltonian, consisting of an infinite number of terms, can be expanded in powers of the fine-structure constant $\alpha$ as
\begin{equation}  
H_{\text{NRQED}} = m + H_0 + H^{(4)} + H^{(5)} + H^{(6)} + \ldots,  
\end{equation}  
where the first term is the electron mass, $H_0$ is the Schrödinger Hamiltonian, and the subsequent terms are labeled by their powers of $\alpha$ (In this article, unless otherwise specified, the superscripts refer to the powers of the fine-structure constant) \cite{IEIDES200163,PhysRevA.100.042510,PhysRevA.110.022813,PhysRevLett.118.233001}. For calculations to a specific precision, one can truncate the effective Hamiltonian using typical values of physical quantities in atomic and molecular systems \cite{PhysRevD.53.4909,PhysRevD.55.7267}:  
\begin{equation}  
p \propto m\alpha, \quad \text{eV} \propto m\alpha^2, \quad e\Vec{E} \propto m\alpha^3, \quad e\Vec{B} \propto m\alpha^4.  
\end{equation}  
leading to  
\begin{equation}  
E_{\text{NRQED}} = m + \alpha^2 E^{(2)} + \alpha^4 E^{(4)} + \alpha^5 E^{(5)} + \alpha^6 E^{(6)} + \ldots.  
\end{equation} 
The expanded terms are, in order, the rest energy term, the non-relativistic energy, the leading-order relativistic correction term, the leading-order quantum electrodynamic (QED) correction term, and the higher-order relativistic and QED correction terms. Despite the great success achieved by this method, the number of terms in the effective Hamiltonian that need to be considered for calculating higher-order corrections, as well as the number of Feynman diagrams that need to be computed, will increase significantly \cite{PhysRevA.72.062102,PhysRevA.107.012810,PhysRevA.103.042809,PhysRevA.106.022816}. This results in a sharp rise in the difficulty of theoretical derivation. Therefore, there is a need to find a method that can automatically derive or calculate all or some of these energy corrections resulting from these effects. A possible approach to achieve this goal may be to directly perform a non-relativistic expansion on the energy that includes QED or relativistic effects.

As the order increases, the singularity of the higher-order effective Hamiltonian also increases gradually. This can be seen from the expansion of the relativistic kinetic energy: $
E_k = \frac{p^2}{2} - \frac{p^4}{8} + \frac{p^6}{16} - \frac{5p^8}{128} + \ldots.$ In the order of $\alpha^6$, the $p^6$ term is divergent for $s$-state hydrogen atom. In the higher orders, the relativistic kinetic energy correction operators of more low angular momentum states will also be divergent. Similar results can be found in other relativistic correction terms. It is crucial and challenging to separate (regularize) and cancel these singularities in the effective Hamiltonian. This determines whether reasonable calculation results can be obtained and is also an important verification of the correctness of the effective theory. Current research focuses on second-order perturbation calculations below the $m\alpha^7$ order \cite{PhysRevA.100.042510,PhysRevA.110.022813,PhysRevLett.118.233001} 
\begin{equation}  
E^{(6)} =\langle H^{(6)}+H^{(4)}\dfrac{Q}{E^{(2)}-H_{0}}H^{(4)}\rangle,  
\end{equation} 
or 
\begin{equation}  
E^{(7)} =\langle H^{(7)}+2 H^{(5)}\dfrac{Q}{E^{(2)}-H_{0}}H^{(4)}\rangle.  
\end{equation} 
Both $H^{(6)/(7)}$ and the second-order perturbation term of $H^{(4)}$ contain divergent terms, which can be separated through appropriate regularization methods to obtain finite and effective operator  $H^{(6)/(7)}_R$  and the second-order perturbation term of $H^{(4)}_R$. However, at the $\alpha^8$ order, which includes the third-order perturbation term, it may occur that the divergent terms cannot be completely separated. At $\alpha^8$ order, the relativistic corrections for the Dirac Coulomb system $\langle H^{(8)}_R\rangle$, $\langle H^{(6)}_R \dfrac{Q}{E^{(2)}-H_{0}}H^{(4)}_R\rangle$ and the third-order perturbation term $\langle H^{(4)}_R \dfrac{Q}{E^{(2)}-H_{0}}H^{(4)}_R\dfrac{Q}{E^{(2)}-H_{0}}H^{(4)}_R\rangle$ are proportional to the harmonic series for the $s$-state of Hydrogen \cite{Zhou_2023}. And Ref.\cite{Zhou_2023} propose combining these terms to eliminate the singularity. It can be inferred that similar problems may also exist in other systems or higher-order correction effects, so it is necessary to study the methods to deal with such singularities. Performing a non-relativistic expansion on the Dirac-Coulomb equation should partially address this issue. Since the eigenvalues and eigenstates of the Dirac-Coulomb equation are finite results, any order of non-relativistic expansion performed on it should also yield finite results. 

In Section II, we derive recursive formulations for the energy and wave function corrections of the Dirac-Coulomb systems at any given order. In Section III, we attempt to address two issues: the non-relativistic expansion of the transition probability amplitude and non-retarded Breit interaction correction. Section IV showcases numerical calculations for the hydrogen atom within the Dirac-Coulomb system up to the $m\alpha^{20}$ order and for the helium atom up to the $m\alpha^{8}$ order. Lastly, we present our conclusions.

\section{The nonrelativistic expansion of the Dirac-Coulomb energy}\label{NRHamiltonian}

\subsection{One-body systems} \label{Onebodysystem}

The Dirac Coulomb Hamiltonian of one-body system is
\begin{equation}
H_{DC}=\beta m+\alpha\cdot p+V=\begin{pmatrix}(m+V)\mathbb{I}_{2\times2}&&\tilde{p}\\\tilde{p}&&(-m+V)\mathbb{I}_{2\times2},\end{pmatrix}
\end{equation}
where $\tilde{p}=\vec{\sigma}\cdot \vec{p}$, $\sigma^i$ is the Pauli matrix. In this work, the abbreviation  $\tilde{F}$ is employed to denote  $\vec{\sigma}\cdot \vec{F}$. For the sake of formula simplification, the identity matrices $\mathbb{I}_{2\times2}$ will be omitted in the subsequent text.

In the Hamiltonian, the operators $m$, $\tilde{p}$, and $V$ correspond to the orders $\alpha^0$, $\alpha^1$, and $\alpha^2$, respectively. The most renowned method for deriving relativistic corrections is the Foldy-Wouthuysen (FW) transformation \cite{PhysRev.78.29,Chen_2023,PhysRevA.110.012808}, which utilizes a unitary transformation to expand the Dirac-Coulomb Hamiltonian in terms of the order of $\alpha^2$. In this article, we employ a non-relativistic expansion approach to solve the Dirac-Coulomb equation  $H_{DC}\psi = E_{DC}\psi$ by expanding the large component $\varphi$ and the small component $\chi$ of the Dirac wavefunction $\psi$, as well as the energy $E$, in powers of $\alpha$. We utilize the following methodology: $\varphi = \sum_{i=1}^{\infty} \varphi^{(2i)}, \quad \chi = \sum_{i=1}^{\infty} \chi^{(2i+1)}, \quad E = m + \sum_{i=1}^{\infty} E_{DC}^{(2i)}$ Here, the superscripts $(2i)$ and $(2i+1)$ represent the respective orders of the expansion in $\alpha$. The iteration equations are

\begin{equation}\begin{aligned}
&\begin{cases}\tilde{p}\chi^{(3)}-(E_{DC}^{(2)}-V)\varphi^{(2)}=0\\\tilde{p}\varphi^{(2)}-2m\chi^{(3)}=0\end{cases}, \\
&\begin{cases}\tilde{p}\chi^{(2n+1)}-(E_{DC}^{(2)}-V)\varphi^{(2n)}=E_{DC}^{(2n)}\varphi^{(2)}+\sum_{i=2}^{n-1}E_{DC}^{(2i)}\varphi^{(2n+2-2i)},\\\tilde{p}\varphi^{(2n)}-2m\chi^{(2n+1)}=(E_{DC}^{(2)}-V)\chi^{(2n-1)}+\sum_{i=2}^{n-1}E_{DC}^{(2i)}\chi^{(2n+1-2i)}\end{cases}& (n>1). 
\end{aligned}\end{equation}

The leading-order equations are
\begin{equation}
\begin{cases}H_{0}\varphi^{(2)}=E_{DC}^{(2)}\varphi^{(2)},
\\\chi^{(3)}=\frac{1}{2m}\tilde{p}\varphi^{(2)}.\end{cases}
\end{equation}
The first line represents the Schrödinger equation, where $H_{0}=\frac{1}{2m}\tilde{p}^2+V$. Using this system of equations, one can obtain the leading-order term of the solution for the Dirac Coulomb equation and then utilize the iterative equation to find approximate solutions of any order.
\begin{equation}\begin{aligned}
\begin{cases} &E_{DC}^{(2n)}=-\frac1{2m}(\varphi^{(2)},\tilde{p} A \chi^{(2n-1)}+\sum_{i=1}^{n-2}E_{DC}^{(2i+2)}\tilde{p}\chi^{(2n-1-2i)}),
\\
&\varphi^{(2n)}=-\frac1{2m} G\tilde{p} A \chi^{(2n-1)}
-\sum_{i=1}^{n-2}E_{DC}^{(2i+2)}
G(\frac1{2m} \tilde{p}\chi^{(2n-1-2i)})+\varphi^{(2n-2i)}),
\\
&\chi^{(2n+1)}=\frac1{2m}(\tilde{p}\varphi^{(2n)}
-\tilde{p} A \chi^{(2n-1)}
-\sum_{i=1}^{n-2}E_{DC}^{(2i+2)}\tilde{p}\chi^{(2n-1-2i)})
.\end{cases}& (n>1) 
\end{aligned}\end{equation}
where $A=E_{DC}^{(2)}-V$, $(\phi,\psi)\equiv \langle\phi |\psi\rangle $ and the nonrelativistic reduced Green operator is $G=\frac{Q}{E_{DC}^{(2)}-H_{0}}$.

The energy corrections below the order of $m\alpha^{8}$ are

\begin{equation}\begin{aligned}
E_{DC}^{(4)}=&\langle H^{(4)}_{R}\rangle \\
E_{DC}^{(6)}=&\langle H^{(6)}_{R}\rangle+\frac{1}{2}E_{DC}{(2)}E_{DC}^{(4)}
+\langle H^{(4)}_{R}G H^{(4)}_{R}\rangle.
\\
E_{DC}^{(8)}=&\left\langle H^{(4)}_{R}G\left(H^{(4)}_{R}-E_{(4)}\right)G H^{(4)}_{R}\right\rangle
+\left\langle H^{(4)}_{R}G H^{(6)}_{R}+ H^{(4)}_{R}G H^{(6)}_{R}\right\rangle
+\left\langle H^{(8)}_{R}\right\rangle\\&
-\frac12E_{(4)}\left\langle A G H^{(4)}_{R}
+H^{(4)}_{R} G A\right\rangle-E_{(4)}^2
+\frac12E_{(2)}E_{(6)}
\end{aligned}\end{equation}
where $H^{(4)}_{R}=-\frac{1}{4}\tilde{p}A\tilde{p}, H^{(6)}_{R}=\frac{1}{8}\tilde{p}A^2\tilde{p}$ and $H^{(8)}_{R}=-\frac1{16}\tilde{p}A^3\tilde{p}$. These results are in agreement with the findings in Ref.\cite{Zhou_2023}. It can be easily seen that operators like $H^{(2n)}_{R}\propto\tilde{p}A^{(n-1)}\tilde{p}$ will appear in higher-order operators. Therefore, higher-order relativistic operators $H^{(2n)}_{R}(n>3)$ will inevitably exhibit singularities. However, during numerical computations, the energy eigenvalues of the Dirac Hamiltonian do not exhibit any divergent results. It can also be seen from the iterative process that each step is merely multiplying by operators such as $\tilde{p},A$ and $G$, which are finite matrices, and no singularities will appear. The numerical computation results afterwards also do not show any divergence. The method for handling singularities is basically consistent with that in Ref.\cite{Zhou_2023} below $\alpha^8$ order and can be extended to higher orders.

\subsection{Few-body systems} \label{Fewbodysystem}

This subsection takes the two-body Dirac-Coulomb case as an example to discuss the non-relativistic expansion of the few-body Dirac-Coulomb equation.

The two-body Dirac-Coulomb equation is given by $H_{DC}\psi = E_{DC}\psi$, where the Hamiltonian is 

\begin{equation}  
\begin{aligned}  
H_{DC} &= \beta_1 m_1 + \alpha_1 \cdot p_1 + \beta_2 m_2 + \alpha_2 \cdot p_2 + V \\  
&=   
\begin{pmatrix}  
m_1 & \tilde{p}_1 \\  
\tilde{p}_1 & -m_1  
\end{pmatrix}  
\otimes \mathbb{I}_2 +  
\mathbb{I}_1 \otimes  
\begin{pmatrix}  
m_2 & \tilde{p}_2 \\  
\tilde{p}_2 & -m_2  
\end{pmatrix} +  
V \mathbb{I}_1 \otimes \mathbb{I}_2.  
\end{aligned}  
\end{equation} 
Here, the symbols $\otimes$ and $\mathbb{I}_n$ denote the tensor product and the $n \times n$ identity matrix, respectively. The two-body wave function is expressed as  
\begin{equation} 
\psi =   
\begin{pmatrix}  
\varphi_1 \\  
\chi_1  
\end{pmatrix}  
\otimes  
\begin{pmatrix}  
\varphi_2 \\  
\chi_2  
\end{pmatrix}.  
\end{equation}  
Introduce the following definition of the components of the wavefunction:
$\Phi=\varphi_1\varphi_2$,
$\Phi_1=\chi_1\varphi_2 $,
$\Phi_2=\varphi_1\chi_2$ and 
$\Phi_{12}=\chi_1\chi_2$.
$\varphi$ and $\chi$ represent the large and small components of the wavefunction, respectively. The two-body Dirac-Coulomb equation can further be decomposed into the following coupled equations:

\begin{equation}  
\left\{\begin{aligned}  
\tilde{p}_a \Phi_a=&(E_{DC}-V-M)\Phi  \\  
\tilde{p}_a \Phi+ \sum_{a\neq b}\tilde{p}_b \Phi_{ab}-2m_a \Phi_a =&(E_{DC}-V-M)\Phi_a   \\
\tilde{p}_a \Phi_b+ \tilde{p}_b \Phi_a-(2m_a+2m_b) \Phi_{ab} =&(E_{DC}-V-M)\Phi_{ab}   \\
\end{aligned}\right.  
\end{equation} 

For three-body or n-body systems, wavefunction components can be introduced similarly to those in two-body systems, as follows: $\Phi=\varphi_1\varphi_2 ...\varphi_n$,
$\Phi_1=\chi_1\varphi_2\varphi_3... \varphi_n$,
$\Phi_2=\varphi_1\chi_2\varphi_3... \varphi_n$,
$\Phi_{12}=\chi_1\chi_2\varphi_3...\varphi_n$,
$\Phi_{23}=\varphi_1\chi_2 \chi_3...\varphi_n$,
$\Phi_{123}=\chi_1\chi_2 \chi_3...\varphi_n$ and other similar terms. The number of coupled equations increases as the number of particles increases. Among them, the first few are also the most important equations, referred to 
\begin{equation}  
\left\{\begin{aligned}  
\tilde{p}_a \Phi_a=&(E_{DC}-V-M)\Phi  \\  
\tilde{p}_a \Phi+ \sum_{a\neq b}\tilde{p}_b \Phi_{ab}-2m_a \Phi_a =&(E_{DC}-V-M)\Phi_a   \\
\tilde{p}_a \Phi_b+ \tilde{p}_b \Phi_a + \sum_{c \neq a b}\tilde{p}_c \Phi_{abc}-(2m_a+2m_b) \Phi_{ab} =&(E_{DC}-V-M)\Phi_{ab}   \\
\tilde{p}_a \Phi_{bc}+ \tilde{p}_b \Phi_{ac}+ \tilde{p}_c \Phi_{ab} + \sum_{d \neq a b c}\tilde{p}_d \Phi_{abcd}-(2m_a+2m_b+2m_c) \Phi_{abc} =&(E_{DC}-V-M)\Phi_{abc}\\
....
\end{aligned}\right.  
\end{equation} 
Here and in the subsequent ellipses, they represent the equations of $\Phi_{abc},\Phi_{abcd}....$

In this subsection, we expand the energy and wave function in the following manner:
\begin{equation}
\left\{\begin{aligned}
E=\left(m_a+m_b\right. & \left.+m_c+\ldots\right)+\sum_i E_{DC}^{(2 i)}\\
\Phi & =\sum_{i=1}^n \Phi^{(2 i)} \\
\Phi_a & =\sum_{i=1} \Phi_a^{(2 i+1)} \\
\Phi_{a b} & =\sum_{i=1} \Phi_{a b}^{(2 i+2)} \\
\Phi_{a b c} & =\sum_{i=1} \Phi_{a b c}^{(2 i+3)} \\
...
\end{aligned}\right.
\end{equation}
By incorporating this result into the Dirac equation, we can obtain the equation that needs to be satisfied by the higher-order energy and wave function.

The leading-order wavefunctions fulfill the following system of equations:
\begin{equation}
\left\{\begin{aligned}
& H_{0} \Phi^{(2)}=E_{DC}^{(2)} \Phi^{(2)} \\
& \Phi_a^{(3)}=\frac{1}{2 m_a} \tilde{p}_a \Phi^{(2)} \\
& \Phi_{a b}^{(4)}=\frac{1}{2 m_a} \frac{1}{2 m_b} \tilde{p}_a \tilde{p}_b \Phi^{(2)} \\
& \Phi_{a b c}^{(5)}=\frac{1}{2 m_a} \frac{1}{2 m_b} \frac{1}{2 m_c} \tilde{p}_a \tilde{p}_b \tilde{p}_c \Phi^{(2)}\\
&...
\end{aligned}\right.
\end{equation}
The first line is the n-body Schrödinger equation, where $H_{0}=\sum_{i}\frac{1}{2m_i}p^2_i+V$.  The simplified recurrence relation of higher-order energy and wave function are:
\begin{equation}
\left\{\begin{aligned}
E_{DC}^{(2 n)}&=\sum_{a \neq b} \frac{1}{2 m_a}\left(\Phi^{(2)}, \tilde{p}_a \tilde{p}_b \Phi_{a b}^{(2 n)}\right)-\sum_a \frac{1}{2 m_a}\left(\tilde{p}_a \Phi^{(2)}, A \Phi_a^{(2 n-1)}+\sum_{i=1}^{n-2} E_{DC}^{(2 i+2)} \Phi_a^{(2 n-2 i-1)}\right), 
\\
\Phi^{(2 n)}&=\sum_{a \neq b} G \frac{\tilde{p}_a \tilde{p}_b}{2 m_a} \Phi_{a b}^{(2 n)}-\sum_a G \frac{\tilde{p}_a}{2 m_a} A \Phi_a^{(2 n-1)}-\sum_{i=1}^{n-2} E_{DC}^{(2 i+2)} G\left(\Phi^{(2 n-2 i)}+\sum_a \frac{\tilde{p}_a}{2 m_a} \Phi_a^{(2 n-2 i-1)}\right), 
\\
\Phi_a^{(2 n+1)}&=\frac{1}{2 m_a}\left(\tilde{p}_a \Phi^{(2 n)}+\sum_{b\neq a}\tilde{p}_b \Phi_{a b}^{(2 n)}-A \Phi_a^{(2 n-1)}-\sum_{i=1}^{n-2} E_{DC}^{(2 i+2)} \Phi_a^{(2 n-2 i-1)}\right),
\\
\Phi_{a b}^{(2 n)}&=\frac{1}{2\left(m_a+m_b\right)}(\tilde{p}_a \Phi_b^{(2 n-1)}+\tilde{p}_b \Phi_a^{(2 n-1)}+\sum_{c \neq a, b} \tilde{p}_c \Phi_{a b c}^{(2 n-1)}-A \Phi_{a b}^{(2 n-2)})
\\
&-\frac{1}{2\left(m_a+m_b\right)}\sum_{i=1}^{n-3} E_{DC}^{(2 i+2)} \Phi_{a b}^{(2 n-2 i-2)}, 
\\
\Phi_{a b c}^{(2 n-1)}&=\frac{1}{2\left(m_a+m_b+m_c\right)}(\tilde{p}_a \Phi_{b c}^{(2 n-2)}+\tilde{p}_b \Phi_{a c}^{(2 n-2)}+\tilde{p}_c \Phi_{a b}^{(2 n-2)}+\sum_{d \neq a b, c} \tilde{p}_d \Phi_{a b c d}^{(2 n-2)}-A \Phi_{a b c}^{(2 n-3)})
\\
&-\frac{1}{2\left(m_a+m_b+m_c\right)}\sum_{i=1}^{n-4} E_{DC}^{(2 i+2)} \Phi_{a b c}^{(2 n-2 i-3)},
\\
&...
\end{aligned}\right.
\end{equation}
where $A=E_{DC}^{(2)}-V$, $(\phi,\psi)\equiv \langle\phi |\psi\rangle $ and the nonrelativistic reduced Green operator is $G=\frac{Q}{E_{DC}^{(2)}-H_{0}}$.

Based on the energy recurrence relation, the leading-order relativistic energy correction operator can be derived ($\alpha^4$ order). It is
\begin{equation}
H^{(4)}_R=
\sum_{a<b}\frac{2m_a+2m_b}{(2m_a)^2(2m_b)^2}\tilde{p}_a^2\tilde{p}_b^2
-\sum_a\frac1{(2m_a)^2}\tilde{p}_a A \tilde{p}_a,
\end{equation}
and the $\alpha^6$ order energy correction is 
\begin{equation}
E_{DC}^{(6)}=\langle H_R^{(6)}+H_R^{(4)}GH_R^{(4)} \rangle,
\end{equation} 
where $H_R^{(6)}$ is $\alpha^6$ order  energy correction operator
\begin{equation}\begin{aligned}  
H^{(6)}_R=&
\sum_{a<b}\left(\dfrac{(1+m_a^2/m_b^2)p_{a}^{4}p_{b}^{2}+(1+m_b^2/m_a^2)p_{a}^{2}p_{b}^{4}}{(2m_a)^2 (2m_b)^2 (2m_a+2m_b)}
-\frac{\tilde{p}_a \tilde{p}_b A \tilde{p}_b \tilde{p}_a}{(2m_a)^2 (2m_b)^2}
\right)
\\&
-\sum_{a<b}\left(\frac1{(2m_a)^3 (2m_b)}\{\tilde{p}_a A \tilde{p}_a, p_{b}^{2}\}
+\frac1{(2m_a) (2m_b)^3}\{\tilde{p}_b A \tilde{p}_b, p_{a}^{2}\}\right)
\\&
+\sum_a\left(
\frac{\tilde{p}_a A^{2} \tilde{p}_a}{(2m_a)^3}
-E_{DC}^{(4)} \dfrac{p_{a}^{2}}{(2m_{a})^{2}}\right)
\\&
+\sum_{a<b<c}(m_a+m_b+m_c+\dfrac{2 m_b m_c}{m_a}+\dfrac{2 m_a m_c}{m_b}+\dfrac{2 m_a m_b}{m_c})\dfrac{p_{a}^{2}p_{b}^{2}p_{c}^{2}}{32m_a^2m_b^2m_c^2}.
\end{aligned}\end{equation} 
Higher-order correction operators can be obtained through iterative methods.

From the recurrence formula expanded by the wavefunction $\Phi$. They are 
\begin{equation}
\left\{\begin{aligned}
&\Phi^{(4)}=GH_R^{(4)}\Phi^{(2)}\\
&\Phi^{(6)}=(GH_R^{(6)}+G(H_R^{(4)}-E_{DC}^{(4)})G H_R^{(4)})\Phi^{(2)}\\
...
\end{aligned}\right.
\end{equation}
it can be seen that each order of the expansion $\Phi^{(2 n)}$ corresponds to higher-order relativistic corrections of the non-relativistic wavefunction $\Phi^{(2)}$.

For a two-electron system, the above results can be expressed in the following form. The $\alpha^4$ order Hamiltonian is 
\begin{equation}H^{(4)}_R=\frac{1}{4m^3}
\tilde{p}_1^2\tilde{p}_2^2-\frac1{4m^2}\tilde{p}_1 A \tilde{p}_1-\frac1{4m^2}\tilde{p}_2 A \tilde{p}_2,
\end{equation}
It is equivalent to the Hamiltonian obtained through the FW transformation of Dirac-Coulomb Hamiltonian at $\alpha^4$ order,
\begin{equation}
H^{(4)}_{DC-FW}=-\dfrac{p^{4}_{1}}{8m^3}-\dfrac{p^{4}_{2}}{8m^3}
-\dfrac{1}{8m^2}[\tilde{p}_1,[\tilde{p}_1,V ] ]
-\dfrac{1}{8m^2}[\tilde{p}_2,[\tilde{p}_2,V ] ].
\end{equation}
Compared to the leading term in relativity \citep{PhysRevA.100.042510}, the contribution from the transverse photon exchange is missing here. We will complete this term in Section III. The $\alpha^6$ order Hamiltonian is
\begin{equation}\begin{aligned}  
H^{(6)}_R=&-\dfrac{1}{4m^2}E_{DC}^{(4)}(p_{1}^{2}+p_{2}^{2})
+\dfrac{1}{64m^5}\{p_{1}^{2}p_{2}^{2},(p_{1}^{2}+p_{2}^{2})\}
+\frac1{8m^3}\tilde{p}_1 A^{2} \tilde{p}_1
+\frac1{8m^3}\tilde{p}_2 A^{2} \tilde{p}_2
\\&
-\frac1{16m^4}\{\tilde{p}_1 A \tilde{p}_1, p_{2}^{2}\}
-\frac1{16m^4}\{\tilde{p}_2 A \tilde{p}_2, p_{1}^{2}\}
-\frac1{16m^4}\tilde{p}_1 \tilde{p}_2 A \tilde{p}_2 \tilde{p}_1
\end{aligned}\end{equation} 
These results are equivalent to the findings in Ref.\cite{Zhou_2023}.

\section{Non-relativistic expansion of transition probability amplitude and non-retarded Breit interaction }\label{one-photon exchange interaction}

The non-relativistic expansion of the wavefunction can be used to derive higher-order relativistic corrections for other physical effects. This section will expand on the transition amplitude and energy correction of non-retarded Breit interaction.

Firstly, the Dirac wavefunction obtained from the expansion of the normalized non-relativistic wavefunction is not normalized. The square of the modulus of the Dirac wavefunction also requires a non-relativistic expansion. The result of the transition amplitude and single-photon exchange interaction needs to be divided by this expansion result or its normalization factor
\begin{equation}
\begin{aligned}
\langle\Psi|\Psi\rangle^{(2n)}=
\sum_{n=0} \langle\Psi|\Psi\rangle^{(2n)},
\end{aligned}
\end{equation}
where 
\begin{equation}
\begin{aligned}
\langle\Psi|\Psi\rangle^{(2n)}=
& \sum_{i=0}^{n}
\langle\Phi^{(2+2i)}|\Phi^{(2+2n-2i)}\rangle
+\sum_{a}\sum_{i=0}^{n-1}
\langle\Phi_a^{(3+2i)}| \Phi_a^{(1+2n-2i)}\rangle
\\&
+\sum_{a<b}\sum_{i=0}^{n-2}
\langle\Phi_{ab}^{(4+2i)}| \Phi_{ab}^{(2n-2i)}\rangle+\sum_{a<b<c}\sum_{i=0}^{n-3}
\langle\Phi_{abc}^{(6+2i)}| \Phi_{abc}^{(2n-2i)}\rangle+....
\end{aligned}
\end{equation}
The leading term is $\langle\Psi|\Psi\rangle^{(0)}=1$ and the second order term is $\langle\Psi|\Psi\rangle^{(2)}=\sum_a\langle\Psi^{(2)}| \frac{\tilde{p}_a^2}{ m_a^2} |\Psi^{(2)}\rangle$.

The transition probability amplitude obtained from the unnormalized wavefunction is  
\begin{equation}\begin{aligned}
\langle\Psi| 
\vec{\epsilon}\cdot\vec{\alpha_a}
 e^{i k \cdot \vec{r}_a}
\left|\Psi^{\prime}\right\rangle=&
\langle\Phi| 
\vec{\epsilon}\cdot\vec{\sigma_a}
e^{i \vec{k} \cdot \vec{r}_a}\left|\Phi_a^{\prime}
\right\rangle+
\left\langle\Phi_a\right| 
\vec{\epsilon}\cdot\vec{\sigma_a}
 e^{i \vec{k} \cdot \vec{r}_a}
\left|\Phi^{\prime}\right\rangle+
\left\langle\Phi_b\right| 
\vec{\epsilon}\cdot\vec{\sigma_a}
 e^{i \vec{k} \cdot \vec{r}_a}
\left|\Phi_{a b}^{\prime}\right\rangle
\\&+
\left\langle\Phi_{a b}\right| 
\vec{\epsilon}\cdot\vec{\sigma_a}
 e^{i \vec{k} \cdot \vec{r}_a}
\left|\Phi_b^{\prime}\right\rangle+...
\\
=&
\sum_{n=0}^{\infty}\langle\Psi| 
\vec{\epsilon}\cdot\vec{\alpha_a} 
e^{i \vec{k} \cdot \vec{r}_a}
\left|\Psi^{\prime}\right\rangle^{(2n+1)},
\end{aligned}
\end{equation}
where the non-relativistic expansion term is
\begin{equation}
\begin{aligned}
\langle\Psi| \vec{\epsilon}\cdot\vec{\alpha_a}
e^{i \vec{k} \cdot \vec{r}_a}
\left|\Psi^{\prime}\right\rangle^{(2n+1)}
=&\sum_{i=0}^{n}
\left(\left\langle\Phi^{(2 i)}\right| 
\vec{\epsilon}\cdot\vec{\sigma_a}
 e^{i \vec{k} \cdot \vec{r}_a}
\left|\Phi_a^{\prime(2 n-2 i+3)}\right\rangle+
\left\langle\Phi_a^{(2 n-2 i+3)}\right| 
\vec{\epsilon}\cdot\vec{\sigma_a}
 e^{i \vec{k} \cdot \vec{r}_a}
\left|\Phi^{\prime(2 i)}\right\rangle
\right) \\
& +\sum_{b\neq a}\sum_{i=0}^{n-1}
\left(\left\langle\Phi_b^{(2 i+3)}\right| 
\vec{\epsilon}\cdot\vec{\sigma_a}
 e^{i \vec{k} \cdot \vec{r}_a}
\left|\Phi_{a b}^{\prime(2 n-2 i+4)}\right\rangle+\left\langle\Phi_{a b}^{(2 n-2 i+4)}\right| 
\vec{\epsilon}\cdot\vec{\sigma_a}
 e^{i \vec{k} \cdot \vec{r}_a}\left|\Phi_b^{\prime(2 i+3)}\right\rangle \right) 
 \\&+...
\end{aligned}
\end{equation}

The leading order transition probability amplitude is 
\begin{equation}
\begin{aligned}
J^{(1)}=&\langle\Psi| 
\vec{\epsilon}\cdot\vec{\alpha_a}
 e^{i \vec{k} \cdot \vec{r}_a}
\left|\Psi^{\prime}\right\rangle^{(1)}
\\
=&
\left\langle\Phi^{(2)}\right| 
\vec{\epsilon}\cdot\vec{\sigma_a}
 e^{i \vec{k} \cdot \vec{r}_a}
\left|\Phi_a^{\prime(3)}\right\rangle+
\left\langle\Phi_a^{(3)}\right| 
\vec{\epsilon}\cdot\vec{\sigma_a}
 e^{i \vec{k} \cdot \vec{r}_a}
\left|\Phi^{(2)}\right\rangle
\\
=&
\left\langle\Phi^{(2)}\right|
\left\{
\vec{\epsilon}\cdot\vec{\sigma_a}
 e^{i \vec{k} \cdot \vec{r}_a},
 \frac{\tilde{p}_a}{2 m_a}\right\}
\left|\Phi^{(2)}\right\rangle,
\end{aligned}
\end{equation}
It is evident that this is equivalent to the scattering probability amplitude induced by The FW Hamiltonian for the leading term of photon-electron coupling $(\tilde{p}-e\tilde{A})^{2}/(2m)$, where $\tilde{A}$ is the inner product of the electromagnetic vector potential and the Pauli matrice \cite{Zhou_2023}.
and the relativistic corrections to  transition probability amplitude is 
\begin{equation}
\begin{aligned}
J^{(3)}=\langle\Psi| 
 \vec{\epsilon}\cdot\vec{\alpha_a}
 e^{i \vec{k} \cdot \vec{r}_a}
 \left|\Psi^{\prime}\right\rangle^{(3)}
 -\dfrac{1}{2}\langle\Psi| 
\vec{\epsilon}\cdot\vec{\alpha_a}
 e^{i \vec{k} \cdot \vec{r}_a}
\left|\Psi^{\prime}\right\rangle^{(1)}
(\langle\Psi|\Psi\rangle^{(2)}
+\langle\Psi^{\prime}|\Psi^{\prime}\rangle^{(2)}),
\end{aligned}
\end{equation}
where
\begin{equation}
\begin{aligned}
\langle\Psi| 
 \vec{\epsilon}\cdot\vec{\alpha_a}
 e^{i \vec{k} \cdot \vec{r}_a}
 \left|\Psi^{\prime}\right\rangle^{(3)}=&
 \left\langle\Phi^{(4)}\right|
 \left\{\vec{\epsilon}\cdot\vec{\alpha_a}
  e^{i \vec{k} \cdot \vec{r}_a}, 
  \frac{1}{2 m_a} \tilde{p}_a\right\}
 \left|\Phi^{\prime(2)}\right\rangle+
 \left\langle\Phi^{(2)}\right|
 \left\{\frac{1}{2 m_a} \tilde{p}_a, 
 \vec{\epsilon}\cdot\vec{\alpha_a}
  e^{i \bar{k} \cdot \vec{r}_a}\right\}\left|\Phi^{\prime(4)}\right\rangle 
 \\
& -\frac{1}{\left(2 m_a\right)^2}
\left\langle\Phi^{(2)}\right| 
\vec{\epsilon}\cdot\vec{\alpha_a}
 e^{i \vec{k} \cdot \vec{r}_a} 
A^{\prime} \tilde{p}_a
\left|\Phi^{\prime(2)}\right\rangle
-\frac{1}{\left(2 m_a\right)^2}
\left\langle\Phi^{(2)}\right| 
\tilde{p}_a A 
\vec{\epsilon}\cdot\vec{\alpha_a}
e^{i \vec{k} \cdot \vec{r}_a}
\left|\Phi^{\prime(2)}\right\rangle \\
& +\frac{1}{\left(2 m_a\right)^2}\sum_{b\neq a}
\left(\frac{1}{2 m_b}+\frac{1}{2 m_a}\right)
\left\langle\Phi^{(2)}\right|
\left\{\tilde{p}_a, 
\vec{\epsilon}\cdot\vec{\alpha_a} 
e^{i \vec{k} \cdot \vec{r}_a}\right\} \tilde{p}_b^2\left|\Phi^{\prime(2)}\right\rangle.
\end{aligned}
\end{equation}
Since $\Phi^{(4)}=G H_{R}^{(4)}\Phi^{(2)}$, the first two terms represent the first-order perturbation of the Dirac-Coulomb Hamiltonian to the scattering probability amplitude. The presence of terms involving $p$ cubed and terms that include the Coulomb interaction  $A$ (where $A=E_{DC}^{(2)}-V$) in the other terms indicates that they are due to the NRQED Hamiltonian of secondary photon-electron coupling. Specifically, these additional terms can be interpreted as arising from the single photon-electron coupling term in the relativistic corrections, as well as terms that encompass the coupling between Coulomb photons, transverse photon, and electrons. Higher-order terms should also be higher-order perturbations of these effects, but among them, the transverse photon field $A$ only appears once. Similar results can be found in the relativistic expansion terms of other effects as well.

By performing a non-relativistic expansion on the wavefunction in the above equation using the method described previously, we can also obtain the various orders of contributions of transverse photon exchange with non-retardation approximation. Using the non-retarded Breit potential
$V^T_{ab}=\frac{-q_{a}q_{b}}{4\pi r_{a b}}
\left(\alpha_a \cdot \alpha_b+\frac{\left(\alpha_a \cdot r_{a b}\right)\left(\alpha_b \cdot r_{a b}\right)}
{r_{a b}^2}\right)$,
one can obtain the relativistic correction form of  transverse photon exchange with non-retardation approximation
\begin{equation}
E_{1 pt}=\sum_{a\neq b}\frac{\langle\Psi|V^T_{ab}|\Psi\rangle}{\langle\Psi|\Psi\rangle}.
\end{equation}
Its non-relativistic expansion of the numerator part is
\begin{equation}
\begin{aligned}
\langle\Psi|V^T_{ab}|\Psi\rangle= \sum_{n=0} \langle\Psi|V^T_{ab}|\Psi\rangle^{(4+2n)}.
\end{aligned}
\end{equation}
The right-hand side of the two-body parts are
\begin{equation}
\begin{aligned}
(\langle\Psi|V_T|\Psi\rangle)^{(4+2n)}_{2-body}=
& \sum_{a\neq b}\sum_{i=0}^{n}\left(
\langle\Phi^{(2+2i)}| V^t_{ab} |\Phi_{ab}^{(4+2n-2i)}\rangle
+\langle\Phi_{a}^{(3+2i)}| V^t_{ab} |\Phi_{b}^{(3+2n-2i)}\rangle
\right.\\&
\left.
+\langle\Phi_{b}^{(3+2i)}| V^t_{ab} |\Phi_{a}^{(3+2n-2i)}\rangle
+\langle\Phi_{ab}^{(4+2i)}| V^t_{ab} |\Phi^{(2+2n-2i)}\rangle\right),
\end{aligned}
\end{equation}
and 3-body parts are
\begin{equation}
\begin{aligned}
(\langle\Psi|V_T|\Psi\rangle)^{(6+2n)}_{3-body}=
& \sum_{c\neq a,b} \sum_{i=0}^{n}\left(
\langle\Phi^{(3+2i)}_c| V^t_{ab} |\Phi_{abc}^{(5+2n-2i)}\rangle
+\langle\Phi_{ac}^{(4+2i)}| V^t_{ab} |\Phi_{bc}^{(4+2n-2i)}\rangle
\right.\\&
\left.
+\langle\Phi_{bc}^{(4+2i)}| V^t_{ab} |\Phi_{ac}^{(4+2n-2i)}\rangle
+\langle\Phi_{abc}^{(5+2i)}| V^t_{ab} |\Phi^{(3+2n-2i)}_c\rangle\right),
\end{aligned}
\end{equation}
where 
$V^t_{ab}=\frac{-q_{a}q_{b}}{4\pi r_{a b}}\left(\sigma_a \cdot \sigma_b+\frac{\left(\sigma_a \cdot r_{a b}\right)\left(\sigma_b \cdot r_{a b}\right)}{r_{a b}^2}\right)$.
The results for other few-body components can be obtained by analogy to the above findings, the leading term of n-body parts is $m\alpha^{2n}$.

From this derivation, we can obtain the energy corrections of transverse photon exchange with non-retardation approximation to any given order. The results up to $m\alpha^8$ order are
\begin{equation}
\begin{aligned}
E_{1pt}^{(4)}=&(\langle\Psi|V_T|\Psi\rangle)^{(4)},
\\
E_{1pt}^{(6)}=&(\langle\Psi|V_T|\Psi\rangle)^{(6)}
-(\langle\Psi|V_T|\Psi\rangle)^{(4)}\langle\Psi|\Psi\rangle^{(2)},
\\
E_{1pt}^{(8)}=&(\langle\Psi|V_T|\Psi\rangle)^{(8)}
-(\langle\Psi|V_T|\Psi\rangle)^{(6)}\langle\Psi|\Psi\rangle^{(2)}
-(\langle\Psi|V_T|\Psi\rangle)^{(4)}
(\langle\Psi|\Psi\rangle^{(4)}+(\langle\Psi|\Psi\rangle^{(2)})^2).
\end{aligned}
\end{equation}
The results of these expansion calculations should all be finite, as neither the potential energy function $V^t_{ab}$ nor the high-order relativistic corrections to wave functions obtained via the numerical iterative method described in Section II possess any singularities. Similarly to the previous analysis of higher-order scattering probability amplitudes, here we have contributions that include the perturbation of the Dirac-Coulomb Hamiltonian due to the exchange of transverse photon, as well as relativistic corrections to the exchange of transverse photon. Additionally, there are corrections from terms involving the coupling of transverse photons with multiple Coulomb photons and electrons of NRQED Hamiltonian.

The leading term can be expanded as 
$E_{1pt}^{(4)}=\langle\Phi^{(2)}|V_{1pt}|\Phi^{(2)}\rangle$, where the effective potential is
\begin{equation}
\begin{aligned}
V_{1pt}=\sum_{a\neg b}\{\dfrac{\tilde{p}_a}{m_a},
\{\dfrac{\tilde{p}_b}{m_b}, V^t_{ab}\}\}. 
\end{aligned}
\end{equation}
At the $\alpha^4$ order, The total contribution of $H^{(4)}_{R}+V_{1pt}$ is equivalent to the relativistic correction operator for two-electron system,
\begin{equation}
H^{(4)}_{Rel}=-\dfrac{p^{4}_1}{8m^3}-\dfrac{p^{4}_2}{8m^3}
-\dfrac{1}{8m^2}[\tilde{p}_1,[\tilde{p}_1,V ] ]
-\dfrac{1}{8m^2}[\tilde{p}_2,[\tilde{p}_2,V ] ]
+\{\dfrac{\tilde{p}_1}{m},
\{\dfrac{\tilde{p}_2}{m}, V^t_{12}\}\}.
\end{equation}
Substituting $\tilde{p}_i=\vec{\sigma}_i \cdot \vec{p}_i (i=1,2)$ and expanding, it will yield the result found in Ref.\citep{PhysRevA.100.042510}. Among them, the first two terms represent the relativistic kinetic energy corrections, the third and fourth terms represent the corrections due to Coulomb photon exchange interactions, and the last term represents the correction due to transverse photon exchange interactions.

The total relativistic energy is $E_{DC}+E_{1pt}$ can be expanded in terms of $\alpha^2$ as follows:
\begin{equation}
E_{Rel}=E_{DC}+E_{1pt}=E_{DC}^{(2)}+\sum_{i=2}(E_{DC}^{(2i)}+E_{1pt}^{(2i)}).
\end{equation}
It encompasses all contributions from relativistic kinetic energy, the exchange of Coulomb photons, and exchange of a single transverse photon with non-retarded approximation.

\section{Calculation method and Numerical result}\label{Calculation}

\subsection{Hydrogen} \label{Hydrogen}

For a single-electron system, we will perform calculations using the hydrogen atom system as an example and compare the results with analytical solutions. The hydrogen atom's radial and angular wave functions are expanded using Slater basis functions and spherical harmonics spinors,  
\begin{equation}  
\varphi = r^{i}e^{-\alpha r}\chi_{\kappa m}(\theta,\phi)  
\end{equation}  
where $\alpha$ is a non-linear parameter, and the spherical harmonics spinor is defined as  
\begin{equation}\begin{aligned}  
\chi_{\kappa m}(\theta,\phi) &\equiv \begin{pmatrix}1\\0\end{pmatrix}Y_{l m-\frac{1}{2}}(\theta,\phi)\langle lm,\frac{1}{2}\frac{1}{2}| jm\rangle \\  
&\quad + \begin{pmatrix}0\\1\end{pmatrix}Y_{l m+\frac{1}{2}}(\theta,\phi)\langle lm,\frac{1}{2}-\frac{1}{2}| jm\rangle  
\end{aligned}\end{equation}  
where the relation between the total angular momentum $j$, orbital angular momentum $l$, and the quantum number $\kappa$ is given by $\kappa = -1-l$ if $j = l + \frac{1}{2}$, and $\kappa = l$ if $j = l - \frac{1}{2}$. The result of applying $\tilde{p}$ to this basis function is given by
\begin{equation}
\tilde{p}(g(r)\chi_{\kappa m}) = i\left(g'(r) + \frac{\kappa + 1}{r}g(r)\right)\chi_{-\kappa m}
\end{equation}

For numerical calculations, we first use the variational method to obtain the non-relativistic energy and wave function. Then, utilizing the above equation and the recurrence relations from Section II, we can derive higher-order relativistic energy corrections, which are presented in Table I-IX. By selecting appropriate nonlinear parameters, the non-relativistic energies and wave functions of these states can be obtained with precise results. Therefore, only the relativistic correction parts are listed in the table. For specific states, the convergence rate of high-order relativistic corrections to accurate values is slower than that of low-order ones. For example, when up to 200 basis functions are used, the leading-order correction of the ground state can reach 7 digits, while the $\alpha^{20}$ order correction only has 2 digits. This is precisely caused by iterative calculation. For the relativistic correction of the same order, it can be seen that the convergence speed of the results of low angular momentum states is slower than that of high angular momentum states. For example, for the $\alpha^{20}$ order relativistic correction, the results of numerical calculations from $1s_{1/2}$ to $3d_{3/2}$ can increase from 1 digit of effective number to 5 digits.

\subsection{Helium} \label{ Helium}
 
For the helium atom, we adopt the Slater basis set with J-J coupling for our calculations. The form of the basis functions is represented as 
\begin{equation}  
\varphi = r^{i}_1 r^{j}_2 e^{-\alpha r_1-\beta r_1}
C^{JM}_{j_{1}m_1 j_{2} m_2}
\chi_{\kappa_1 m_1}(\theta_1,\phi_1)  
\chi_{\kappa_2 m_2}(\theta_2,\phi_2),
\end{equation}  
where $C^{JM}_{j_{1}m_1 j_{2} m_2}$ stands for the Clebsch-Gordan Coefficients, $\chi$ is the spherical harmonic spinor, $j_a=|\kappa_a|-1/2$ and $J$ denote the angular momentum quantum number of the electron and the total angular momentum quantum number, respectively, and $\alpha,\beta$ are the nonlinear parameter. 

In the ground state of helium, the total angular momentum is $0$, and the kappa quantum numbers for the two electrons are both $-1$. Since the $ \tilde{p}$ operator reverses the sign of the kappa quantum number, when calculating the relativistic corrections for the ground state of helium, we choose the set of angular kappa quantum number pairs as $(-1,-1),(-1,1), (1,-1)$, and $(1,1)$. For each partial wave, distinct nonlinear parameters are employed, which in this paper are $(1.42,2.91), (2.60,4.10), (4.60,2.61)$, and $(4.16,4.91)$, respectively. Finally, we consider the cases where the sum of the indices $i$ and $j$ is less than or equal to $\Omega$. Table X shows the ground state energy of helium atoms up to the $m\alpha^8$ order.
 
\section{Discussion and Conclusion} \label{Discussion and Conclusion}

Through the method of non-relativistic perturbation expansion of the Dirac equation, we have derived iterative equations for calculating high-order relativistic corrections of Dirac-Coulomb energy. It can be observed that within the order of $\alpha^8$, the relativistic correction operators corresponding to the single-electron system are consistent with the results in the literature \cite{Zhou_2023}. Using this method, the high-order correction operators of the hydrogen atom can be rapidly converged to the exact value up to the order of $\alpha^{20}$, and the convergence speed increases with the increase of angular momentum. This is precisely because the higher-order terms exhibit higher singularity for states with lower angular momentum. For intermediate states, we need to consider the behavior of the wave function near the nucleus, thus requiring higher nonlinear parameters. For two-electron systems, we have calculated the relativistic corrections up to the order of $\alpha^8$ for the ground state of the Dirac helium atom based on the Slater basis with JJ coupling. However, the convergence speed is not ideal. This may be due to insufficient accuracy of the wave function and inadequate research on the main factors affecting the contribution of intermediate states. Therefore, there is a need to further develop optimization methods for non-relativistic perturbation expansion.

Upon delving into the relativistic correction of the single-electron Dirac-Coulomb Hamiltonian up to the $\alpha^8$ order, we find that it is necessary to decompose the higher-order operators with singularities into matrix products of lower-order operators to cancel the singularities in the higher-order perturbed Hamiltonian. This correction is the same as the result of the regulariation of the FW-transformed Hamiltonian \cite{Zhou_2023}. We therefore speculate that in the traditional NRQED approach, the use of FW transformation to obtain equivalent operators and then derive the normalized energy correction operators is also likely to require the same method to deal with the singularities of order $\alpha^8$ . In contrast, the non-relativistic expansion method can avoid cumbersome additional regularization processes.

In addition, the method of non-relativistic expansion of relativistic wave functions can also be utilized to expand the probability amplitudes or energy corrections of various physical effects. Our results for the non-relativistic expansion of the transition amplitude associated with the exchange of transverse photons agree with those of NRQED in the leading term. In higher-order corrections, our findings include contributions from all Coulomb photons as well as those from single transverse photon. Notably, the $\alpha^4$ order correction to the Dirac Coulomb energy and the non-retarded Breit interaction correspond precisely to the $\alpha^4$ order relativistic correction. Higher-order expansion terms contribute at even powers of alpha, and theoretically, each order should yield finite results. Therefore, we consider these higher-order corrections as higher-order forms of relativistic corrections. Finally, compared to the traditional NRQED method, the non-relativistic expansion method may significantly reduce the extensive workload of theoretical calculations. 

\textbf{ACKNOWLEDGMENTS}

We acknowledges helpful conversations with Yongbo Tang and Liming Wang. This work was supported by the National Natural Science Foundation of China (Nos.12074295 and 12304271).

\bibliography{1.bib}

\newpage

\begin{table}[htbp] 
\caption{The relativistic corrections to energy of ground state $1 s_{1/2}$ of hydrogen atom up to $\alpha^{10}$. One of nonlinear parameters is $1$, and the non-relativistic energy of the ground state is the exact value. The first column is the number of basis functions, and the remaining columns are the relativistic higher-order corrections $E_{DC}^{(n)}$ divided by $\alpha^n$ in atomic units.}
\begin{tabular}{lllllll} 
\hline  Num~~ & $E_{DC}^{(4)}$~~ & $E_{DC}^{(6)}$~~ & $E_{DC}^{(8)}$~~ & $E_{DC}^{(10)}$~~ \\ \hline\hline
40 & $-0.12476192$~~ & $-0.061378940$~~ & $-0.036300173$~~ & $-0.022534454$~~ \\[3pt]
60 & $-0.12495454$~~ & $-0.062237485$~~ & $-0.038291606$~~ & $-0.025765552$~~ \\[3pt] 
80 & $-0.12499125$~~ & $-0.062444615$~~ & $-0.038878503$~~ & $-0.026913790$~~ \\[3pt] 
100 & $-0.12499614$~~ & $-0.062473346$~~ & $-0.038967166$~~ & $-0.027106083$~~\\[3pt] 
120 & $-0.12499806$~~ & $-0.062483997$~~ & $-0.039001099$~~ & $-0.027182975$~~\\[3pt] 
140 & $-0.12499900$~~ & $-0.062492384$~~ & $-0.039032604$~~ & $-0.027262145$~~\\[3pt] 
160 & $-0.12499935$~~ & $-0.062494195$~~ & $-0.039038463$~~ & $-0.027276053$~~\\[3pt] 
180 & $-0.12499961$~~ & $-0.062496450$~~ & $-0.039047282$~~ & $-0.027299440$~~\\[3pt]
200 & $-0.12499976$~~ & $-0.062497655$~~ & $-0.039052204$~~ & $-0.027312876$~~\\[3pt]
$\infty$ & $-0.12499996(20)$~~ & $-0.0624990(25)$~~ & $-0.039058(6)$ & $-0.027331(18)$~~ \\[3pt]
Exact & $-0.125$~~ & $-0.0625$~~ & $-0.0390625$~~ & $-0.02734375$~~ \\[3pt] 
\hline   
\end{tabular}
\end{table}

\begin{table}[htbp] 
\caption{The relativistic corrections to energy of ground state $1 s_{1/2}$ of hydrogen atom up to $\alpha^{20}$. One of nonlinear parameters is $1$, and the non-relativistic energy of the ground state is the exact value. The first column is the number of basis functions, and the remaining columns are the relativistic higher-order corrections $E_{DC}^{(n)}$ divided by $\alpha^n$ in atomic units.}
\begin{tabular}{llllll} 
\hline  Num~~ & $E_{DC}^{(12)}$~~ & $E_{DC}^{(14)}$~~ & $E_{DC}^{(16)}$~~ & $E_{DC}^{(18)}$~~ & $E_{DC}^{(20)}$~~\\ \hline\hline
40 & $-0.013784875$~~& $-0.008008730$~~ & $-0.0042786466$  & $-0.0019971832$ & $-0.00070803194$ \\[3pt]
60 & $-0.017938961$~~& $-0.012540280$~~ & $-0.0086501968$  & $-0.0058235487$ & $-0.0037934943$\\[3pt] 
80 & $-0.019709329$~~& $-0.014852225$~~ & $-0.011324094$  & $-0.0086443727$ & $-0.0065632798$\\[3pt] 
100 & $-0.020039993$~~& $-0.015334242$~~& $-0.011945540$  & $-0.0093737817$ & $-0.0073588997$\\[3pt] 
120 & $-0.020179343$~~& $-0.015547937$~~& $-0.012235453$  & $-0.0097305107$ & $-0.0077666329$\\[3pt] 
140 & $-0.020332568$~~& $-0.015796282$~~& $-0.012587677$  & $-0.010182987$ & $-0.0083035160$\\[3pt] 
160 & $-0.020359585$~~& $-0.015840948$~~& $-0.012653341$  & $-0.010270367$ & $-0.0084115905$\\[3pt] 
180 & $-0.020407667$~~& $-0.015923696$~~& $-0.012777912$  & $-0.010439760$ & $-0.0086240277$\\[3pt]
200 & $-0.020436455$~~& $-0.015974656$~~& $-0.012857682$  & $-0.010550580$ & $-0.0087689024$\\[3pt]
$\infty$ & $-0.02048(5)$~~& $-0.01606(9)$~~& $-0.01300(15)$  & $-0.01076(21)$ & $-0.00908(32)$\\[3pt]
Exact  & $-0.0205078...$~~& $-0.0161132...$~~& $-0.01309...$  & $-0.01091...$ & $-0.00927...$\\[3pt] 
\hline   
\end{tabular}
\end{table}

\begin{table}[htbp] 
\caption{The relativistic corrections to energy of $2 p_{1/2}$ state of hydrogen atom up to $\alpha^{10}$. One of nonlinear parameters is $1/2$, and the non-relativistic energy  is the exact value. The first column is the number of basis functions, and the remaining columns are the relativistic higher-order corrections $E_{DC}^{(n)}$ divided by $\alpha^n$ in atomic units.}
\begin{tabular}{lllllll} 
\hline  Num~~ & $E_{DC}^{(4)}$~~ & $E_{DC}^{(6)}$~~ & $E_{DC}^{(8)}$~~ & $E_{DC}^{(10)}$~~ \\ \hline\hline
40 & $-0.03906270796$~~ & $-0.020499387$~~ & $-0.013043798$~~ & $-0.0091274220$~~ \\[3pt]
60 & $-0.03906253594$~~ & $-0.020507314$~~ & $-0.013088358$~~ & $-0.0092596985$~~ \\[3pt] 
80 & $-0.03906251393$~~ & $-0.020507684$~~ & $-0.013090955$~~ & $-0.0092690557$~~ \\[3pt] 
100 & $-0.03906250672$~~ & $-0.020507699$~~ & $-0.013091165$~~ & $-0.0092699773$~~\\[3pt] 
120 & $-0.03906250363$~~ & $-0.020507699$~~ & $-0.013091283$~~ & $-0.0092705334$~~\\[3pt] 
140 & $-0.03906250213$~~ & $-0.020507736$~~ & $-0.013091549$~~ & $-0.0092715285$~~\\[3pt] 
160 & $-0.03906250133$~~ & $-0.020507761$~~ & $-0.013091714$~~ & $-0.0092721582$~~\\[3pt] 
180 & $-0.03906250086$~~ & $-0.020507785$~~ & $-0.013091854$~~ & $-0.0092727224$~~\\[3pt]
200 & $-0.03906250059$~~ & $-0.020507799$~~ & $-0.013091937$~~ & $-0.0092730628$~~\\[3pt]
$\infty$ & $-0.03906250022(37)$~~ & $-0.020507816(17)$~~ & $-0.01309205(12)$ & $-0.0092736(5)$~~ \\[3pt]
Exact & $-0.0390625$~~ & $-0.0205078125$~~ & $-0.01309204...$~~ & $-0.0092735...$~~ \\[3pt] 
\hline   
\end{tabular}
\end{table}

\begin{table}[htbp] 
\caption{The relativistic corrections to energy of $2 p_{1/2}$ state of hydrogen atom up to $\alpha^{20}$. One of nonlinear parameters is $1/2$, and the non-relativistic energy is the exact value. The first column is the number of basis functions, and the remaining columns are the relativistic higher-order corrections $E_{DC}^{(n)}$ divided by $\alpha^n$ in atomic units.}
\begin{tabular}{llllll} 
\hline  Num~~ & $E_{DC}^{(12)}$~~ & $E_{DC}^{(14)}$~~ & $E_{DC}^{(16)}$~~ & $E_{DC}^{(18)}$~~ & $E_{DC}^{(20)}$~~\\ \hline\hline
40 & $-0.0066935000$~~& $-0.0049933524$~~ & $-0.0037159598$  & $-0.0027228663$ & $-0.0019464794$ \\[3pt]
60 & $-0.0069714096$~~& $-0.0054588295$~~ & $-0.0043795491$  & $-0.0035621765$ & $-0.0029150982$\\[3pt] 
80 & $-0.0069950304$~~& $-0.0055060262$~~ & $-0.0044591333$  & $-0.0036802610$ & $-0.0030737411$\\[3pt] 
100 & $-0.0069977250$~~& $-0.0055121056$~~& $-0.0044705420$  & $-0.0036988336$ & $-0.0031008553$\\[3pt] 
120 & $-0.0069995321$~~& $-0.0055162190$~~& $-0.0044786792$  & $-0.0037119926$ & $-0.0031211144$\\[3pt] 
140 & $-0.0070022784$~~& $-0.0055220101$~~& $-0.0044896404$  & $-0.0037287897$ & $-0.0031478096$\\[3pt] 
160 & $-0.0070039999$~~& $-0.0055257375$~~& $-0.0044967430$  & $-0.0037398915$ & $-0.0031660746$\\[3pt] 
180 & $-0.0070054888$~~& $-0.0055292089$~~& $-0.0045029191$  & $-0.0037512400$ & $-0.0031807386$\\[3pt]
200 & $-0.0070063991$~~& $-0.0055314040$~~& $-0.0045068791$  & $-0.0037587717$ & $-0.0031906038$\\[3pt]
$\infty$ & $-0.0070078(14)$~~& $-0.0055352(21)$~~& $-0.004514(7)$  & $-0.003774(15)$ & $-0.003211(20)$\\[3pt]
Exact  & $-0.0070078...$~~& $-0.0055350...$~~& $-0.004514...$  & $-0.003773....$ & $-0.003214...$\\[3pt] 
\hline   
\end{tabular}
\end{table}

\begin{table}[htbp] 
\caption{The relativistic corrections to energy of $2 p_{3/2}$ state of hydrogen atom up to $\alpha^{10}$. One of nonlinear parameters is $1/2$, and the non-relativistic energy  is the exact value. The first column is the number of basis functions, and the remaining columns are the relativistic higher-order corrections $E_{DC}^{(n)}$ divided by $\alpha^n$ in atomic units.}
\begin{tabular}{lllllll} 
\hline  Num~~ & $E_{DC}^{(4)}$~~ & $E_{DC}^{(6)}$~~ & $E_{DC}^{(8)}$~~ & $E_{DC}^{(10)}$~~ \\ \hline\hline
40 & $-0.00781231280892$~~ & $-0.00097638901144$~~ & $-0.00015247681736$~~ & $-0.00002663048289$~~ \\[3pt]
60 & $-0.00781249405689$~~ & $-0.00097655706282$~~ & $-0.00015258555908$~~ & $-0.00002669988222$~~ \\[3pt] 
80 & $-0.00781249964348$~~ & $-0.00097656206840$~~ & $-0.00015258747769$~~ & $-0.00002670252051$~~ \\[3pt] 
100 & $-0.00781249993756$~~ & $-0.00097656243252$~~ & $-0.00015258782877$~~ & $-0.00002670282349$~~\\[3pt] 
120 & $-0.00781249998486$~~ & $-0.00097656248354$~~ & $-0.00015258787494$~~ & $-0.00002670286570$~~\\[3pt] 
140 & $-0.00781249999549$~~ & $-0.00097656249482$~~ & $-0.00015258788584$~~ & $-0.00002670287554$~~\\[3pt] 
160 & $-0.00781249999860$~~ & $-0.00097656249814$~~ & $-0.00015258788859$~~ & $-0.00002670287855$~~\\[3pt] 
180 & $-0.00781249999943$~~ & $-0.00097656249925$~~ & $-0.00015258788982$~~ & $-0.00002670287992$~~\\[3pt]
200 & $-0.00781249999975$~~ & $-0.00097656249967$~~ & $-0.00015258789024$~~ & $-0.00002670288043$~~\\[3pt]
$\infty$ & $-0.00781249999994(19)$~~ & $-0.00097656249991(26)$~~ & $-0.00015258789046(22)$ & $-0.00002670288072(29)$~~ \\[3pt]
Exact & $-0.0078125$~~ & $-0.0009765625$~~ & $-0.000152587890625$~~ & $-0.00002670288085...$~~ \\[3pt] 
\hline   
\end{tabular}
\end{table}

\begin{table}[htbp] 
\caption{The relativistic corrections to energy of $2 p_{3/2}$ state of hydrogen atom up to $\alpha^{20}$. One of nonlinear parameters is $1/2$, and the non-relativistic energy is the exact value. The first column is the number of basis functions, and the remaining columns are the relativistic higher-order corrections $E_{DC}^{(n)}$ divided by $\alpha^n$ in atomic units.}
\begin{tabular}{llllll} 
\hline  Num~~ & $E_{DC}^{(12)}\times 10^{6}$~~ & $E_{DC}^{(14)}\times 10^{7}$~~ & $E_{DC}^{(16)}\times 10^{7}$~~ & $E_{DC}^{(18)}\times 10^{8}$~~ & $E_{DC}^{(20)}\times 10^{9}$~~\\ \hline\hline
40 & $-4.9638913$~~& $-9.6171393$~~ & $-1.902270$  & $-3.789231$ & $-7.52196$ \\[3pt]
60 & $-5.0057904$~~& $-9.8166822$~~ & $-2.000282$  & $-4.048845$ & $-9.66547$\\[3pt] 
80 & $-5.0065207$~~& $-9.8330229$~~ & $-1.996692$  & $-4.156763$ & $-8.82032$\\[3pt] 
100 & $-5.0067444$~~& $-9.8344486$~~& $-1.9974919$  & $-4.160775$ & $-8.83858$\\[3pt] 
120 & $-5.0067776$~~& $-9.8346751$~~& $-1.9976281$  & $-4.161508$ & $-8.84215$\\[3pt] 
140 & $-5.0067856$~~& $-9.8347305$~~& $-1.9976633$  & $-4.161702$ & $-8.84316$\\[3pt] 
160 & $-5.0067881$~~& $-9.8347495$~~& $-1.9976752$  & $-4.161775$ & $-8.84352$\\[3pt] 
180 & $-5.0067893$~~& $-9.8347592$~~& $-1.9976818$  & $-4.161815$ & $-8.84374$\\[3pt]
200 & $-5.0067898$~~& $-9.8347630$~~& $-1.9976845$  & $-4.161832$ & $-8.84383$\\[3pt]
$\infty$ & $-5.0067900(13)$~~& $-9.8347654(24)$~~& $-1.9976861(16)$  & $-4.161843(12)$ & $-8.84390(7)$\\[3pt]
Exact  & $-5.0067901...$~~& $-9.8347663...$~~& $-1.9976869...$  & $-4.161847...$ & $-8.84392...$\\[3pt] 
\hline   
\end{tabular}
\end{table}

\begin{table}[htbp] 
\caption{The relativistic corrections to energy of $3 d_{3/2}$ state of hydrogen atom up to $\alpha^{10}$. One of nonlinear parameters is $1/3$, and the non-relativistic energy  is the exact value. The first column is the number of basis functions, and the remaining columns are the relativistic higher-order corrections $E_{DC}^{(n)}$ divided by $\alpha^n$ in atomic units.}
\begin{tabular}{lllllll} 
\hline  Num~~ & $E_{DC}^{(4)}$~~ & $E_{DC}^{(6)}$~~ & $E_{DC}^{(8)}$~~ & $E_{DC}^{(10)}$~~ \\ \hline\hline
40 &  $-0.004629629653863317$~~ & $-0.00062156682569$~~ & $-0.00010031503960$~~ & $-0.00001787898828$~~ \\[3pt]
60 &  $-0.004629629631309066$~~ & $-0.00062157062499$~~ & $-0.00010032049187$~~ & $-0.00001788284463$~~ \\[3pt] 
80 &  $-0.004629629629843316$~~ & $-0.00062157064133$~~ & $-0.00010032053794$~~ & $-0.00001788291220$~~ \\[3pt] 
100 & $-0.004629629629659949$~~ & $-0.00062157064348$~~ & $-0.00010032054510$~~ & $-0.00001788292406$~~\\[3pt] 
120 & $-0.004629629629634524$~~ & $-0.00062157064420$~~ & $-0.00010032054758$~~ & $-0.00001788292834$~~\\[3pt] 
140 & $-0.004629629629630471$~~ & $-0.00062157064443$~~ & $-0.00010032054841$~~ & $-0.00001788292982$~~\\[3pt] 
160 & $-0.004629629629629707$~~ & $-0.00062157064456$~~ & $-0.00010032054885$~~ & $-0.00001788293061$~~\\[3pt] 
180 & $-0.004629629629629646$~~ & $-0.00062157064464$~~ & $-0.00010032054915$~~ & $-0.00001788293117$~~\\[3pt]
200 & $-0.004629629629629633$~~ & $-0.00062157064468$~~ & $-0.00010032054931$~~ & $-0.00001788293148$~~\\[3pt]
$\infty$& $-0.004629629629629630(3)$~~ & $-0.00062157064473(5)$~~ & $-0.00010032054952(21)$ & $-0.00001788293190(42)$~~ \\[3pt]
Exact   & $-0.004629629629629629...$~~ & $-0.00062157064471...$~~ & $-0.00010032054945...$~~ & $-0.00001788293177...$~~ \\[3pt] 
\hline   
\end{tabular}
\end{table}

\begin{table}[htbp] 
\caption{The relativistic corrections to energy of $3 d_{3/2}$ state of hydrogen atom up to $\alpha^{20}$. One of nonlinear parameters is $1/3$, and the non-relativistic energy is the exact value. The first column is the number of basis functions, and the remaining columns are the relativistic higher-order corrections $E_{DC}^{(n)}$ divided by $\alpha^n$ in atomic units.}
\begin{tabular}{llllll} 
\hline  Num~~ & $E_{DC}^{(12)}\times 10^{6}$~~ & $E_{DC}^{(14)}\times 10^{7}$~~ & $E_{DC}^{(16)}\times 10^{7}$~~ & $E_{DC}^{(18)}\times 10^{8}$~~ & $E_{DC}^{(20)}\times 10^{9}$~~\\ \hline\hline
40  & $-3.3911926$~~& $-6.712722$~~ & $-1.37050$  & $-2.863225$ & $-6.08491$ \\[3pt]
60  & $-3.3930329$~~& $-6.719745$~~ & $-1.37294$  & $-2.871825$ & $-6.11641$\\[3pt] 
80  & $-3.3931020$~~& $-6.720295$~~ & $-1.37331$  & $-2.873868$ & $-6.12660$\\[3pt] 
100 & $-3.3931153$~~& $-6.720410$~~& $-1.373388$  & $-2.874358$ & $-6.12919$\\[3pt] 
120 & $-3.3931204$~~& $-6.720455$~~& $-1.373421$  & $-2.874564$ & $-6.13032$\\[3pt] 
140 & $-3.3931222$~~& $-6.720472$~~& $-1.373433$  & $-2.874645$ & $-6.13077$\\[3pt] 
160 & $-3.3931231$~~& $-6.720481$~~& $-1.373440$  & $-2.874690$ & $-6.13103$\\[3pt] 
180 & $-3.3931238$~~& $-6.720487$~~& $-1.373445$  & $-2.874725$ & $-6.13124$\\[3pt]
200 & $-3.3931242$~~& $-6.720491$~~& $-1.373449$  & $-2.874747$ & $-6.13137$\\[3pt]
$\infty$ 
    & $-3.3931248(6)$~~& $-6.720497(6)$~~& $-1.373454(5)$  & $-2.874784(37)$ & $ -6.13161(24)$\\[3pt]
Exact  & $-3.3931246...$~~& $-6.720495...$~~& $-1.373452...$  & $-2.874772...$ & $-6.13153...$\\[3pt] 
\hline   
\end{tabular}
\end{table}

\begin{table}[htbp] 
\caption{The relativistic corrections to energy of state of hydrogen atom up to $\alpha^{10}$. The number of basis functions is $40$, and uncertainty is given by the difference between the exact value and the calculated value. The relativistic higher-order corrections $E_{DC}^{(n)}$ are divided by $\alpha^n$ in atomic units. The nonlinear parameters are $1/n$ and $10/n$.}
\begin{tabular}{lllllll} 
\hline  State~~ & $E_{DC}^{(4)}$~~ & $E_{DC}^{(6)}$~~ & $E_{DC}^{(8)}$~~ & $E_{DC}^{(10)}$~~ \\ \hline\hline
$3d_{5/2}$ &  $-0.0015432097(2)$~~ & $-0.0000857336(4)$~~ & $-5.9535(3)\times 10^{-6}$~~ & $-4.629(2)\times 10^{-7}$~~ \\[3pt]
$4f_{5/2}$ &  $-0.00113932283(9)$~~ & $-0.0000672515(3)$~~ & $-4.7912(2)\times 10^{-6}$~~ & $-3.777(2)\times 10^{-7}$~~ \\[3pt] 
$4f_{7/2}$ &  $-0.0004882812495(6)$~~ & $-0.0000152587889(8)$~~ & $-5.960460(5)\times 10^{-7}$~~ & $-2.60768(2)\times 10^{-8}$~~ \\[3pt] 
$5g_{7/2}$ &  $-0.0003999999996(5)$~~ & $-0.0000131249993(7)$~~ & $-5.226558(5)\times 10^{-7}$~~ & $-2.30965(2)\times 10^{-8}$~~ \\[3pt]
$5g_{9/2}$ &  $-0.00019999999997(3)$~~ & $-3.99999998(2)\times 10^{-6}$~~ & $-9.9999996(5)\times 10^{-8}$~~ & $-2.7999991(9)\times 10^{-9}$~~ \\[3pt] 
$6h_{9/2}$ &  $-0.00017361111108(3)$~~ & $-3.61153976(2)\times 10^{-6}$~~ & $-9.1617661(6)\times 10^{-8}$~~ & $-2.584282(2)\times 10^{-9}$~~ \\[3pt] 
$6h_{11/2}$ &  $-0.00009645061727(2)$~~ & $-1.339591900(7)\times 10^{-6}$~~ & $-2.3256803(1)\times 10^{-8}$~~ & $-4.522155(1)\times 10^{-10}$~~ \\[3pt] 
$7i_{11/2}$ &  $-0.00008676940161(3)$~~ & $-1.24448177(1)\times 10^{-6}$~~ & $-2.1852018(2)\times 10^{-8}$~~ & $-4.272775(2)\times 10^{-10}$~~ \\[3pt] 
$7i_{13/2}$ &  $-0.00005206164096(2)$~~ & $-5.31241229(6)\times 10^{-7}$~~ & $-6.7760354(8)\times 10^{-9}$~~ & $-9.680046(6)\times 10^{-11}$~~ \\[3pt] 
$8j_{13/2}$ &  $-0.00004795619416(4)$~~ & $-5.0255577(1)\times 10^{-7}$~~ & $-6.468083(2)\times 10^{-9}$~~ & $-9.28027(2)\times 10^{-11}$~~ \\[3pt] 
$8j_{15/2}$ &  $-0.00003051757811(2))$~~ & $-2.38418574(5)\times 10^{-7}$~~ & $-2.3283058(6)\times 10^{-9}$~~ & $-2.54658(5)\times 10^{-11}$~~ \\[3pt] 
$9k_{15/2}$ &  $-0.00002857796064(4)$~~ & $-2.2831864(1)\times 10^{-7}$~~ & $-2.245968(2)\times 10^{-9}$~~ & $-2.4650178(1)\times 10^{-11}$~~ \\[3pt]
$9k_{17/2}$ &  $-0.00001905197377 (2)$~~ & $-1.17604772 (4)\times 10^{-7}$~~ & $-9.074438 (5)\times 10^{-10}$~~ & $-7.84208(3)\times 10^{-12}$~~ \\[3pt]
\hline   
\end{tabular}
\end{table}

\begin{table}[htbp] 
\caption{The relativistic corrections to ground state energy of Dirac-Coulomb Hamiltonian of Helium up to $\alpha^{8}$. The relativistic higher-order corrections $E_{DC}^{(n)}$ are divided by $\alpha^n$ in atomic units.}
\begin{tabular}{lllll} 
\hline  $\Omega$~~&~~ $E_{DC}^{(2)}$  & ~~$E_{DC}^{(4)}$ & ~~$E_{DC}^{(6)}$ & ~~$E_{DC}^{(8)}$  \\ \hline\hline
1 & $-2.863360$~~  & $-2.1824$~~  & $-0.3774$~~  & $~~1.881$~~  \\[3pt] 
2 & $-2.884090$~~  & $-2.5001$~~  & $-1.700$~~  & $-0.350$~~    \\[3pt] 
3 & $-2.885501$~~  & $-2.4451$~~  & $-2.646$~~  & $-2.205$~~  \\[3pt] 
4 & $-2.886034$~~  & $-2.4537$~~  & $-3.202$~~  & $-3.660$~~  \\[3pt] 
5 & $-2.886231$~~  & $-2.4637$~~  & $-3.522$~~  & $-4.812$~~  \\[3pt] 
6 & $-2.886336$~~  & $-2.4724$~~  & $-3.742$~~  & $-5.667$~~  \\[3pt] 
7 & $-2.886384$~~  & $-2.4801$~~  & $-3.892$~~  & $-6.361$~~  \\[3pt] 
8 & $-2.886414$~~  & $-2.4855$~~  & $-4.003$~~  & $-6.929$~~  \\[3pt] 
9 & $-2.886431$~~  & $-2.4900$~~  & $-4.088$~~  & $-7.391$~~  \\[3pt] 
10 & $-2.886443$~~  & $-2.4932$~~  & $-4.156$~~  & $-7.776$~~  \\[3pt] 
11 & $-2.886452$~~  & $-2.4959$~~  & $-4.211$~~  & $-8.101$~~  \\[3pt] 
12 & $-2.886458$~~  & $-2.4979$~~  & $-4.256$~~  & $-8.376$~~  \\[3pt] 
13 & $-2.886464$~~  & $-2.4995$~~  & $-4.293$~~  & $-8.610$~~  \\[3pt] 
14 & $-2.886468$~~  & $-2.5008$~~  & $-4.325$~~  & $-8.804$~~  \\[3pt] 
15 & $-2.886471$~~  & $-2.5019$~~  & $-4.352$~~  & $-8.976$~~  \\[3pt] 
16 & $-2.886473$~~  & $-2.5029$~~  & $-4.376$~~  & $-9.128$~~  \\[3pt] 
17 & $-2.886475$~~  & $-2.5036$~~  & $-4.396$~~  & $-9.269$~~  \\[3pt] 
18 & $-2.886477$~~  & $-2.5043$~~  & $-4.414$~~  & $-9.391$~~  \\[3pt] 
19 & $-2.886478$~~  & $-2.5049$~~  & $-4.431$~~  & $-9.501$~~  \\[3pt] 
20 & $-2.886480$~~  & $-2.5054$~~  & $-4.445$~~  & $-9.590$~~  \\[3pt] 
$\infty$ & $-2.886486(6)$~~  & $-2.5085(31)$~~  & $-4.55(11)$~~  & $-9.97(38)$~~  \\[3pt] 
\hline   
\end{tabular}
\end{table}

\newpage

\newpage

\end{document}